\documentclass[amssymb,amsmath,aps,secnumarabic,floatfix,twocolumn,pre,showpacs,showkeys,superscriptaddress,balancelastpage]{revtex4}

\usepackage{graphicx}
\usepackage{xspace}
\usepackage{subfigure}
\usepackage{longtable}
\usepackage{multirow}
\usepackage{color}
\usepackage{psfrag}
\usepackage{pstricks}

\newcommand{\latin}[1]{{\it #1}}
\newcommand{\ie}{\latin{i.e.}\@\xspace}
\newcommand{\eg}{\latin{e.g.}\@\xspace}

\makeatletter
\newcommand{\supmarker}[1]{{\@ifempty{#1}{}{\text{(#1)}}}}
\makeatother
\newcommand{\pdf}[2]{\mathcal{P}^{(#1)}\left(#2\right)}

\newcommand{\elabel}[1]{\label{eq:#1}}

\newcommand{\Eref}[1]{Eq.~(\ref{eq:#1})}

\newcommand{\flabel}[1]{\label{fig:#1}}

\newcommand{\BFref}[1]{Figure~\ref{fig:#1}}
\newcommand{\Fref}[1]{Fig.~\ref{fig:#1}}

\newcommand{\tlabel}[1]{\label{table:#1}}

\newcommand{\Tref}[1]{Table~\ref{table:#1}}
\newcommand{\Treft}[2]{Tables~\ref{table:#1} and \ref{table:#2}}

\newcommand{\GC}{\mathcal{G}}

\newcommand{\ave}[1]{\left\langle #1 \right\rangle}

\newcommand{\numact}{N_a}

\newcommand{\momrat}[2]{g^{(#1)}_{#2}}

\newcommand{\qoflow}[1]{[#1]}
\newcommand{\muTilde}{\tilde{\mu}}

\newcommand{\strong}[1]{}

\expandafter\ifx\csname package@font\endcsname\relax\else
  \expandafter\expandafter
  \expandafter\usepackage
  \expandafter\expandafter
  \expandafter{\csname package@font\endcsname}
\fi

\begin{document}
\title{Abelian Manna model in three dimensions and below}
\date{\today}

\author{Hoai Nguyen Huynh}
\email{n.huynh10@imperial.ac.uk}
\homepage{http://www3.ntu.edu.sg/home2008/hu0004en/}
\affiliation{Division of Physics and Applied Physics,
School of Physical and Mathematical Sciences,
Nanyang Technological University,
21 Nanyang Link, Singapore 637371, Singapore}

\author{Gunnar Pruessner}
\email{g.pruessner@imperial.ac.uk}
\homepage{http://www.ma.ic.ac.uk/~pruess/}
\affiliation{Department of Mathematics,
Imperial College London,
180 Queen's Gate,
London SW7 2BZ, United Kingdom}

\begin{abstract}
The Abelian Manna model of self-organized criticality is studied on various
three-dimensional and fractal lattices. The exponents for avalanche size,
duration and area distribution of the model are obtained by using a
high-accuracy moment analysis. Together with earlier results on
lower-dimensional lattices, the present results reinforce the notion of
universality below the upper critical dimension and allow us to determine the
coefficients of an $\epsilon$-expansion. By rescaling the critical exponents by
the lattice dimension and incorporating the random walker dimension , a
remarkable relation is observed, satisfied by both regular and fractal lattices.
\end{abstract}



\maketitle

\section{Introduction}

Critical phenomena play an important role in our understanding of complex
systems in nature. One of the key features of critical systems is the notion of
universality \cite{Stanley:1999}. It suggests common underlying mechanisms of
seemingly different phenomena and also allows us to study complicated natural
systems through analysis of simple (numerical) models. A huge number of
numerical models have been proposed to study different features of critical
systems. Traditionally, those systems require external fine tuning of control
parameter to critical point. However, many models exhibit the features of the
crititcal state without the need of tuning a control parameter, which is known
as self-organized criticality \cite{Bak.etal:1987}. While models of traditional
critical phenomena have been very well studied both analytically and
numerically, and important results including exact ones have been obtained
\cite[\eg][]{Syozi:1951}, little success has been achieved for self-organized
critical phenomena. One example of such models is the Manna model
\cite{Manna:1991} which has so far defied any attempt for an analytical
approach, but has been studied extensively numerically. Yet the question of
universality seems to have been overlooked in the literature in a very
fundamental point: Does the Manna model display the same critical behaviour on
different lattices of the same dimension? Recent extensive numerical studies
\cite{Huynh.etal:2011} of the Abelian version of the Manna model
\cite{Manna:1991,Dhar:1999a,Dhar:1999b} on various lattices in integer
dimensions $1$ and $2$ provide very strong support for the model's universal
behaviour. The situation, however, is more complicated in non-integer dimensions
\cite{Huynh.etal:2010}. Thus, it would be of great interest to see if the
results in fractal dimensions can be reconciled with those in integer dimensions
in a systematic manner.

Our motivation for this study is to provide a complete numerical picture of the
Manna model. Three main results are reported: Firstly, we confirm
universality in three dimensions, \ie the independence of critical exponents and moment ratios from the detailed
structure of the underlying lattice. This allows us, secondly, to firmly
estimate the coefficients of an $\epsilon$-expansion for the exponents. Thirdly,
we identify a general scaling relation unifying critical behaviour on regular
and fractal lattices.

\section{Model and observables}
The Abelian Manna model \cite{Manna:1991,Dhar:1999a,Dhar:1999b,Huynh.etal:2011}
is defined on a lattice of dimension $d$ with $N$ sites and linear extent $L$,
where $N\propto L^d$ asymptotically. Each site $i$ has a local degree of freedom
$z_i\ge0$ which can be thought of as the number of particles residing at that
site. If $z_i>1$, the site is said to be active or unstable, otherwise it is
stable. The system evolves by driving and subsequently fully relaxing it.
Driving: The system is ``charged'' by picking a site $j$ randomly and uniformly
and increasing $z_j$ by $1$. Relaxation: An unstable site $i$ is picked randomly
and uniformly; its particle number is reduced by $2$ and the particle number of
two of its $q_i$ neighbours, which are chosen independently and at random
(possibly the same one twice), is increased by $1$, thereby possibly rendering
them unstable. Dissipation takes place only when relaxing boundary sites
transfer particles to $q^{(v)}$ virtual neighbours ``outside'' the lattice.
These virtual neighbours cannot topple themselves and are chosen so that the
topology of the finite lattice corresponds to that of an ``offcut'' from an
infinite lattice. Relaxation of unstable sites continues until there are no
unstable sites left. Only then the system is driven again, known as a separation
of time scales. The number of topplings between two driving steps is the size
$s$ of the ``avalanche'' and the number of distinct sites toppling is the area
$a$. The duration $t$ of the avalanche is measured on the microscopic time
scale, which advances in steps of $1/\numact$, where $\numact$ is the
instantaneous number of active sites, mimicking a Poissonian decay of active
sites. The moments of the observables mentioned above are measured in the
stationary state, which is reached after a generously estimated transient. A
number of (asymptotic) key characteristics of the lattices are listed in
\Tref{3Dparameters}, such as the average number of neighbours $\overline{q}$,
the average number of virtual neighbours $\overline{q^{(v)}}$ among sites with
at least one virtual neighbour, and the particle density $\zeta$ in the
stationary state.

The probability densities $\pdf{x}{x,L}$ of the observables, $x\in \{s,a,t\}$,
are expected to display finite size scaling
\begin{equation}
\pdf{x}{x,L} = a_x x^{-\tau_x} \GC_x\left(\frac{x}{b_xL^{D_x}}\right),
\end{equation}
provided that $L\gg1$ and $x\gg x_0$ for some lower cutoff $x_0$.
The metric
factors $a_x$ and $b_x$ are not expected to be universal
\cite{PrivmanHohenbergAharony:1991}, whereas the exponent $D_x$ should only
depend on the dimension and $\tau_x$ on the dimension and the boundary
conditions \cite{Nakanishi.Sneppen:1997}. The universal scaling function $\GC_x$
is characterised below by moment ratios. For $n>\tau_x-1$, the moments scale
asymptotically like $\ave{x^n}\propto L^{\mu_n^{(x)}}$ with $\mu_n^{(x)}=D_x
(1+n-\tau_x)$. For historic reason, the exponents are denoted as $D$ (for
$D_s$), $\tau$ (for $\tau_s$), $z$ (for $D_t$), and $\alpha$ (for $\tau_t$).

Five three-dimensional and five fractal lattices are employed in this study. The
three-dimensional lattices \cite{Ashcroft.Mermin:1976} are built upon the
standard simple cubic (SC) lattice. The body centered cubic (BCC) and face
centered cubic (FCC) lattices are also studied with next nearest neighbour
interactions (BCCN and FCCN, respectively). The total number of sites $N$ of all
five lattices are chosen to be as close as possible to one another. Typically,
six system sizes ranging from $N=181^3$ to $N=1024^3$ are used. A total of
approximately 50,000 CPU hours have been spent on three-dimensional
systems. 

The key features of the fractal lattices are listed in \Tref{FRparameters}. Of
those the lesser known semi-inverse square triadic Koch (SSTK) lattice
\cite{Huynh.Chew:2011,Addison:1997} has Hausdorff dimension $d=\ln5/\ln3$ and is
exemplified in \Fref{FRlattices_SSTK}. Of particular interest is the Sierpinski
tetrahedron (SITE) lattice, which is the three-dimensional version of the
well-known Sierpinski gasket, based on tetrahedra instead of triangles. Its
fractal dimension is $d=2$ and thus allows a direct comparison to regular
two-dimensional lattices. The strongly anisotropic extended Sierpinski gasket
(EXGA, $d=\ln6/\ln2$) is obtained by stacking $L$ copies of a Sierpinski gasket
on top of one another and applying periodic boundary conditions. Finally, the
arrowhead (ARRO, $d=\ln{3}/\ln{2}$) and the crab (CRAB, $d=\ln{3}/\ln{2}$)
lattices are the same as the ones used in \cite{Huynh.etal:2010}. Typically,
four system sizes corresponding to iterations from $6$ to $9$ are used for all
fractal lattices ($5$ to $8$ for SSTK). A total of approximately 75,000 CPU
hours have been spent on fractal lattices.

\begin{figure}
\includegraphics[scale=0.4]{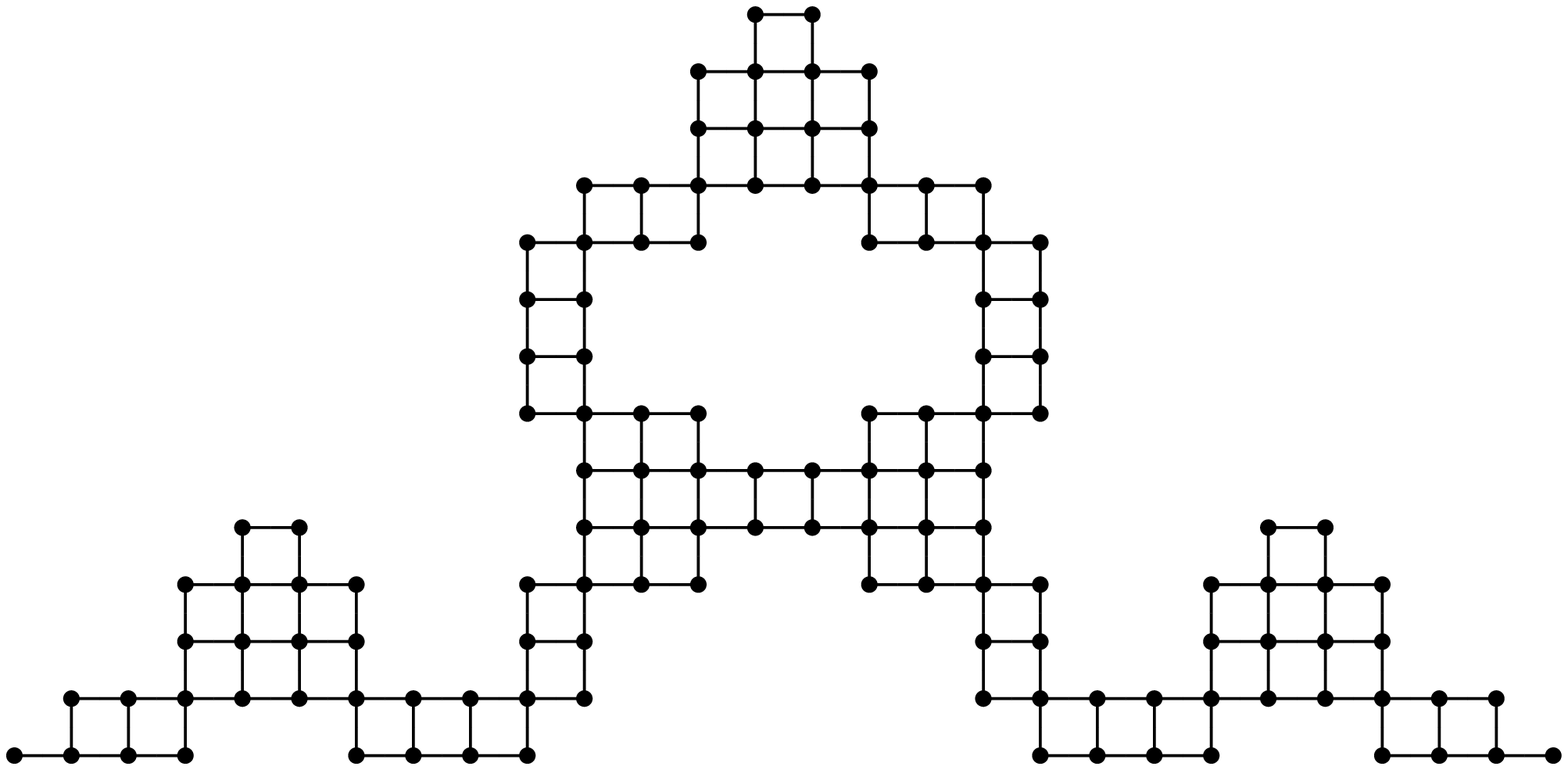}
\caption{\flabel{FRlattices_SSTK}
Semi-inverse square triadic Koch lattice at iteration 3.
}
\end{figure}

\section{Results}
Details of the Monte Carlo simulation, the fitting procedures and the
derivation of the error bars can be found in \cite{Huynh.etal:2011}. In short,
individual moments $\ave{x^n}(L)$ are fitted against the system size $N\propto
L^d$ to obtain the scaling exponent $\mu_n^{(x)}/d$ using a power law with
corrections. For regular three-dimensional lattices, the form of fitting
function used for avalanche size and duration is
\begin{equation}
\ave{x^n}(N) = A_n^{(x)} N^{\mu_n^{(x)}/d} + B_n^{(x)} N^{\mu_n^{(x)}/d-0.25}
\end{equation}
and for avalanche area is
\begin{eqnarray}
\ave{a^n}(N) & = & A_n^{(a)}N^{\mu_n^{(a)}/d}+B_n^{(a)}N^{\mu_n^{(a)}/d-0.25}
\nonumber \\
& + & C_n^{(a)}N^{\mu_n^{(a)}/d-0.5}.
\end{eqnarray}
For fractal lattices we use Eq. (2) in \cite{Huynh.etal:2010}
\begin{eqnarray}
\ave{x^n}(N) & = & A_n^{(x)}N^{\mu_n^{(x)}/d}+B_n^{(x)}N^{\mu_n^{(x)}/d-1}
\nonumber \\
& + & C_n^{(x)}N^{\mu_n^{(x)}/d-2}.
\end{eqnarray}
The estimated scaling exponents $\mu_n^{(x)}$ are then linearly
fitted against
the moment orders $n$ ($n=2,3,4,5$ for all except for $\mu_n^{(s)}$ in three
dimensions with $n=2,3,4$). The slope gives the exponent $D_x$ and the
interception with the abscissa gives the exponent $\tau_x$, except for $\tau_s$
of three-dimensional lattices, whose estimate is obtained by employing the exact
relation $D_s(2-\tau_s)=2$. Although the relative errors are as small as
$3\times 10^{-3}$, we were unable to adjust the fitting scheme as to recover
$\mu_1^{(s)}=2$ within less than $3$ standard deviations.

The quality of all data fitting reported in this work is assessed by the
goodness-of-fit $q$ \cite{Press.etal:2007}, which is considered good if $q>0.1$,
otherwise they are marked by $\qoflow{\cdot}$. \Treft{3Dexponents}{FRexponents}
summarise the estimated critical exponents, all obtained with $q>0.5$. For
regular lattices, our results compare well with the literature
\cite{Ben-Hur.Biham:1996,Luebeck:2000,Pastor-Satorras.Vespignani:2001}
(also \cite{Alava.Munoz:2002,Luebeck.Heger:2003a} for absorbing
state phase transitions), although some
variability and discrepancy is observed in particular for $z$ which may be
explained by the use of slightly different model definitions (and dynamics) by
other authors. A number of scaling relations (see below) are confirmed, such as
$-\Sigma_x=D_x(\tau_x-1)=D_y(\tau_y-1)=-\Sigma_y$ \cite{Luebeck:2000}. Overall
estimates are included in \Tref{3Dexponents}, the correlation of $\Sigma_s$,
$\Sigma_t$ and $\Sigma_a$ is taken into account by multiplying their respective
errors by $\sqrt{3}$. 

\begin{table}
\caption{\tlabel{3Dparameters}Key features of the three-dimensional ($d=3$)
lattices considered in this work. The random walker dimension $d_w$ is $2$ for
every regular lattice \cite{Avraham.Havlin:2000}. The asymptotic site averaged
number of nearest neighbours is $\overline{q}$, with boundary sites having on
average $\overline{q^{(v)}}$ neighbours missing (virtual neighbours
\cite{Huynh.etal:2011}). The stationary particle density, equal to the average
density of (singly) occupied sites, is given by $\zeta$.}
\begin{tabular}{llllll}
\hline\hline
Lattice         & $d$ 	& $d_w$ & $\overline{q}$& $\overline{q^{(v)}}$ 	& $\zeta$                \\
\hline
SC 		& 3 	& 2 	& 6		& 1			& \qoflow{0.622325(1)}	\\
BCC 		& 3 	& 2 	& 8		& 4			& \qoflow{0.600620(2)}	\\
BCCN 		& 3 	& 2 	& 14		& 5			& \qoflow{0.581502(1)}	\\
FCC 		& 3 	& 2 	& 12		& 4			& \qoflow{0.589187(3)}	\\
FCCN 		& 3 	& 2 	& 18		& 5			& \qoflow{0.566307(3)}	\\

\hline\hline
\end{tabular}
\end{table}

The moment ratios \cite{Binder:1981,Huynh.etal:2011} $g_n^{(x)}$ (to leading
order) are independent of the system size and characterise the scaling function
$\GC_x$. The fitting of moment ratios follows a similar procedure as the
avalanche exponents, using
\begin{equation}
g_n^{(x)} = \mathcal{A}_n^{(x)} + \mathcal{B}_n^{(x)} N^{-0.25} +
\mathcal{C}_n^{(x)} N^{-0.5}.
\end{equation}
Together with the avalanche exponents, they provide very strong
support for universality in regular lattices. \Tref{3Dmratio} lists the overall
moment ratios based on five three-dimensional lattices.

Surprisingly (see \cite{Huynh.etal:2011} for the same phenomenon in one and two
dimensions), the \emph{amplitudes} of the leading order of the moments of the
avalanche area seem to be universal. We found
\begin{equation}
\begin{array}{rcl}
\ave{a^1} & = & [0.202(4)]N^{\muTilde_1^{(a)}} \\
\ave{a^2} & = & 0.0151(15)N^{\muTilde_2^{(a)}} \\
\ave{a^3} & = & 0.0027(6)N^{\muTilde_3^{(a)}} \\
\ave{a^4} & = & 0.00055(19)N^{\muTilde_4^{(a)}} \\
\ave{a^5} & = & 0.00012(6)N^{\muTilde_5^{(a)}} \\
\end{array}
\end{equation}
with universal $\muTilde_n^{(a)}=n+1-1.4396(8)$ across the
three-dimensional
lattices introduced above. It is obviously crucial to consider $\ave{a^n}$ as a
function of $N$, as fitting against $L=\lambda N^{1/d}$ leads to different
amplitudes, because $\lambda$ varies from lattice to lattice.

\begin{table}
\caption{\tlabel{FRparameters}
Key features of the fractal lattices studied, as listed in \Tref{3Dparameters}
for the three-dimensional ones. The random walker dimensions are exactly known
or derived (in case of SSTK estimated) using the methods described in
\cite{Huynh.etal:2010}.}
\begin{tabular}{llllll}
\hline\hline
Lattice         & $d$ 		& $d_w$ 	& $\overline{q}$& $\overline{q^{(v)}}$ 	& $\zeta$                \\
\hline
SSTK		& $1.464\ldots$	& 2.552.. 	& 3		& 1			& \qoflow{0.8435(2)}	\\
ARRO		& $1.584\ldots$	& 2.322.. 	& 7/3		& 1			& \qoflow{0.862(2)}	\\
CRAB		& $1.584\ldots$	& 2.578.. 	& 7/3		& 1			& \qoflow{0.8794(6)}	\\
SITE		& $2$ 		& 2.584.. 	& 6		& 3			& \qoflow{0.7427(3)}	\\
EXGA		& $2.584\ldots$	& 2.321.. 	& 6		& 2			& \qoflow{0.65640(8)}	\\

\hline\hline
\end{tabular}
\end{table}

\begin{table*}
\caption{\tlabel{3Dexponents}
Avalanche exponents of five three-dimensional lattices. The estimates for $\tau$
and $D(\tau-1)$ are obtained from $D$ via the exact scaling relation $D(2-\tau)=
2$. Identities $D_a=d$ and $\mu_1^{(s)}=2$ are used to validate the fitting
scheme.}
\begin{tabular}{lllllllllllll}
\hline\hline
Lattice		& $d$ & $d_w$ 	& $D$ 		& $\tau$ 	& $z$ 		& $\alpha$ 	& $D_a$ & $\tau_a$ & $\mu_1^{(s)}$ & $-\Sigma_s$ & $-\Sigma_t$ & $-\Sigma_a$ \\
\hline
SC 		& 3 & 2 	& 3.38(2) 	& 1.408(3) 	& 1.779(7) 	& 1.784(9) 	& 3.04(5) & 1.45(4) & 2.0057(5) & 1.38(2) & 1.395(16)	& 1.36(13) \\
BCC 		& 3 & 2 	& 3.36(2) 	& 1.404(4) 	& 1.777(8) 	& 1.78(1) 	& 2.99(2) & 1.444(18) & 2.0030(5) & 1.36(2) & 1.390(19) & 1.33(6) \\
BCCN 		& 3 & 2 	& 3.38(3) 	& 1.408(4) 	& 1.776(9) 	& 1.783(11) 	& 3.01(3) & 1.44(3) & 2.0041(6) & 1.38(3) & 1.39(2) & 1.32(7) \\
FCC 		& 3 & 2 	& 3.35(4)	& 1.402(8) 	& 1.765(16) 	& 1.78(2) 	& 3.1(2) & 1.48(14) & 2.0035(11) & 1.35(4) & 1.37(4) & 1.5(5) \\
FCCN 		& 3 & 2 	& 3.38(4) 	& 1.408(7) 	& 1.781(14) 	& 1.787(18) 	& 3.00(4) & 1.44(3) & 2.0051(8) & 1.38(4) & 1.40(3) & 1.32(9) \\
\hline
Overall 	& 3 & 2 	& 3.370(11) 	& 1.407(2) 	& 1.777(4) 	& 1.783(5) 	& 3.003(14) & 1.442(12) & 2.0042(3) & \multicolumn{3}{c}{1.380(13)} \\
\hline
\cite{Ben-Hur.Biham:1996}& 3 & 2 & 3.33 	& 1.43 		& 1.8 		& 		& & & & & & \\
\cite{Luebeck:2000}      & 3 & 2 & 3.302(10) 	& 		& 1.713(10) 	& 		& & & & & & \\
\cite{Pastor-Satorras.Vespignani:2001} 
			 & 3 & 2 & 3.36(1) 	& 1.41(1) 	& 1.76(1) 	& 1.78(2) 	& & & & & & \\
\cite{Luebeck.Heger:2003b}& 3 & 2 & 		& 1.41(2) 	& 1.823(23) 	& 1.77(4) 	& & & & & & \\

\hline\hline
\end{tabular}
\end{table*}

\begin{table}
\caption{\tlabel{3Dmratio}
Overall estimates of moment ratios for three-dimensional lattices.
}
\begin{tabular}{lllllll}
\hline\hline
Observable & $x$	& $\momrat{x}{3}$ & $\momrat{x}{4}$ & $\momrat{x}{5}$ & $\momrat{x}{6}$ \\
\hline
%SC\xspace & $s$ & \qoflow{2.366(11)} & \qoflow{7.54(11)} & \qoflow{26.9(9)} & \qoflow{92(7)} \\
%BCC\xspace & $s$ & \qoflow{2.388(12)}	& \qoflow{7.97(13)}	& \qoflow{32.0(11)}	& \qoflow{142(10)} \\
%BCCNNN\xspace & $s$ & \qoflow{2.375(13)}	& \qoflow{7.72(14)}	& \qoflow{29.2(12)}	& 118(10) \\
%FCC\xspace & $s$ & 2.357(26) & 7.6(3) & 29(3) & 120(2) \\
%FCCNNN\xspace & $s$ & 2.38(2) & 7.8(2) & 30.4(17) & 128(14) \\
%\hline
%SC\xspace & $t$ & 4.152(9) & 25.82(15) & 200(2) & 1798(34) \\
%BCC\xspace & $t$ & 4.168(8) & 26.05(14) & 203(2) & 1824(33) \\
%BCCNNN\xspace & $t$ & \qoflow{4.158(9)} & \qoflow{25.87(16)} & 200(2) & 1795(35) \\
%FCC\xspace & $t$ & 4.13(2) & 25.5(4) & 195(5) & 1727(67) \\
%FCCNNN\xspace & $t$ & 4.203(15) & 26.5(3) & 209(4) & 1908(56) \\
%\hline
%SC\xspace & $a$ & 2.335(6) & 7.32(6) & 26.8(4) & \qoflow{108(3)} \\
%BCC\xspace & $a$ & 2.332(7) & \qoflow{7.35(6)} & 27.3(4) & 114(3) \\
%BCCNNN\xspace & $a$ & \qoflow{2.333(7)} & \qoflow{7.34(6)} & \qoflow{27.2(4)} & \qoflow{112(3)} \\
%FCC\xspace & $a$ & 2.311(14) & 7.15(12) & \qoflow{25.9(9)} & \qoflow{103(6)} \\
%FCCNNN\xspace & $a$ & 2.332(11) & 7.318(98) & 26.9(7) & 110(5) \\
%\hline
Size & $s$ & 2.373(16) & 7.76(17) & 30.0(14) & 121(8) \\
Duration & $t$ & \qoflow{4.164(6)} & \qoflow{25.99(9)} & \qoflow{201.4(12)} & 1811(18) \\
Area & $a$ & 2.331(4) & 7.30(5) & 27.1(3) & 113(2) \\
\hline\hline
\end{tabular}
\end{table}

\begin{table*}
\caption{\tlabel{FRexponents}
Avalanche exponents of five fractal lattices. }
\centering
\begin{tabular}{lllllllllllll}
\hline\hline
% Nguyen, I think it makes little sense to quote D(2-\tau) in the table. It is much more interesting to state \mu_1^{(s)}.
Lattice & $d$ 		& $d_w$ 	& $D$ 		& $\tau$ 	& $z$ 		& $\alpha$ 	& $D_a$ 	& $\tau_a$ 	& $\mu_1^{(s)}$		& $-\Sigma_s$ & $-\Sigma_t$ & $-\Sigma_a$ \\
\hline
%SSTK & 1.464\ldots   	& \NHknows 	& 2.94(3) 	& 1.13(2) 	& 1.817(17) 	& 1.21(2) 	& 1.466(5) 	& 1.273(11) 	& \NHreplaces{2.56(7)} 	& 0.37(6) & 0.38(4) & 0.399(17) \\
%ARRO & 1.584\ldots   	& 2.322\ldots 	& 2.7938(19) 	& 1.1731(16) 	& 1.6732(12) 	& 1.2797(17) 	& 1.5847(3)	& 1.2985(6) 	& \NHreplaces{2.310(5)}	& 0.484(5) & 0.468(3) & 0.473(1) \\
%CRAB & 1.584\ldots   	& 2.578\ldots 	& 3.020(5) 	& 1.151(4) 	& 1.837(3) 	& 1.237(4) 	& 1.5847(8) 	& 1.2793(17) 	& \NHreplaces{2.564(12)}& 0.456(11) & 0.435(7) & 0.443(3) \\
%SITE & 2             	& \NHknows 	& 3.232(6) 	& 1.211(4) 	& 1.870(4) 	& 1.357(4) 	& 1.9975(9) 	& 1.3388(14) 	& \NHreplaces{2.549(14)}& 0.682(14) & 0.667(8) & 0.677(3) \\
%EXGA & 2.584\ldots   	& \NHknows 	& 3.352(4) 	& 1.312(3) 	& 1.835(3) 	& 1.581(3) 	& 2.5895(6) 	& 1.3915(8) 	& \NHreplaces{2.306(10)}& 1.0461(98) & 1.066(6) & 1.014(2) \\
% I have edited the data slightly to narrow the table
SSTK & 1.464..   	& 2.552.. 	& 2.94(3) 	& 1.13(2) 	& 1.817(17) 	& 1.21(2) 	& 1.466(5) 	& 1.273(11) 	& 2.551(6) 	& 0.37(6) & 0.38(4) & 0.399(17) \\
ARRO & 1.584..   	& 2.322.. 	& 2.793(2) 	& 1.173(2) 	& 1.673(1) 	& 1.280(2) 	& 1.5847(3)	& 1.2985(6) 	& 2.3103(4)	& 0.484(5) & 0.468(3) & 0.473(1) \\
CRAB & 1.584..   	& 2.578.. 	& 3.020(5) 	& 1.151(4) 	& 1.837(3) 	& 1.237(4) 	& 1.5847(8) 	& 1.279(2) 	& 2.5655(12)	& 0.456(11) & 0.435(7) & 0.443(3) \\
SITE & 2             	& 2.584.. 	& 3.232(6) 	& 1.211(4) 	& 1.870(4) 	& 1.357(4) 	& 1.9975(9) 	& 1.339(2) 	& 2.5533(6)	& 0.682(14) & 0.667(8) & 0.677(3) \\
EXGA & 2.584..   	& 2.321.. 	& 3.352(4) 	& 1.312(3) 	& 1.835(3) 	& 1.581(3) 	& 2.5895(6) 	& 1.3915(8) 	& 2.3000(2)	& 1.046(10) & 1.066(6) & 1.014(2) \\
\hline \hline

\end{tabular}
\end{table*}

\begin{table*}
\caption{\tlabel{ADexponents}
Summary of exponents in all dimensions.
}
\begin{tabular}{llllllllllll}
\hline\hline
Lattice & $d$ 				& $d_w$ 	& $D$ 		& $\tau$ 	& $z$ 		& $\alpha$ 	& $D_a$ 	& $\tau_a$ 	& $\mu_1^{(s)}$ & $-\Sigma$ & Ref. \\
\hline
regular & 1 				& 2 		& 2.253(14) 	& 1.112(6) 	& 1.445(10) 	& 1.18(2) 	& 0.998(3) 	& 1.259(11) 	& 1.996(3) & 0.26(2) & \cite{Huynh.etal:2011} \\
SSTK	& 1.464..\textsuperscript{a} 	& 2.552.. 	& 2.94(3) 	& 1.13(2) 	& 1.817(17) 	& 1.21(2) 	& 1.466(5) 	& 1.273(11) 	& 2.551(6) & 0.40(3) & here\\
% ARRO: d_w=2.322 D=2.792(2)
% CRAB: d_w=2.578 D=3.026(2)
ARRO 	& 1.584..\textsuperscript{b} 	& 2.322.. 	& 2.7938(19) 	& 1.1731(16) 	& 1.6732(12) 	& 1.2797(17) 	& 1.5847(3) 	& 1.2985(6) 	& 2.3103(4) & 0.4730(16) & here+\cite{Huynh.etal:2010} \\
CRAB 	& 1.584..\textsuperscript{b} 	& 2.578.. 	& 3.020(5) 	& 1.151(4) 	& 1.837(3) 	& 1.237(4) 	& 1.5847(8) 	& 1.2793(17) 	& 2.5655(12) & 0.442(4) & here+\cite{Huynh.etal:2010} \\
regular	& 2 				& 2 		& 2.750(6) 	& 1.273(2) 	& 1.532(8) 	& 1.4896(96) 	& 1.995(3) 	& 1.382(3) 	& 1.9993(5) 	& 0.761(13) 	& \cite{Huynh.etal:2011} \\
SITE	& 2\textsuperscript{c} 		& 2.584.. 	& 3.232(6) 	& 1.211(4) 	& 1.870(4) 	& 1.357(4) 	& 1.9975(9) 	& 1.3388(14) 	& 2.5533(6) & 0.676(5) &  \\
EXGA	& 2.584..\textsuperscript{d} 	& 2.321.. 	& 3.352(4) 	& 1.312(3) 	& 1.835(3) 	& 1.581(3) 	& 2.5895(6) 	& 1.3915(8) 	& 2.3000(2) & 1.020(3) & \\
regular	& 3 				& 2 		& 3.370(11) 	& 1.407(2) 	& 1.777(4) 	& 1.783(5) 	& 3.003(14) 	& 1.442(12) 	& 2.0043(3) & 1.380(13) & here \\
regular	& 4\textsuperscript{e} 		& 2 		& 4 		& 1.5 		& 2 		& 2 		& 4 		& 1.5 		& 2 & 2 & \cite{Luebeck:2004} \\
\hline \hline
\multicolumn{9}{l}{\textsuperscript{a}$\ln{5}/\ln{3}$.
\textsuperscript{b}$\ln{3}/\ln{2}$. 
\textsuperscript{c}Fractal lattice.
\textsuperscript{d}$\ln{6}/\ln{2}$, strongly anisotropic.}  \\
\multicolumn{9}{l}{\textsuperscript{e}At upper critical dimension $d_c=4$ \cite{Luebeck.Heger:2003a}, exponents take mean-field value \cite{Luebeck:2004}.}

\end{tabular}
\end{table*} 

\subsection{Regular lattices}
All critical exponents including previous results \cite{Huynh.etal:2011} are
summarised in \Tref{ADexponents}. Firstly, on regular lattices, a relation
between $D_x$, $\tau_x$ and the dimension $d$ can be obtained by fitting
exponents against a proposed function $D_x=f_x(d)$ and $\tau_x=h_x(d)$. With six
exponents six functions are to be determined, which, however, are related by
scaling laws. They are
\begin{equation}\elabel{normal_diffusion}
D(2-\tau)=2
\end{equation}
on regular lattices (exact \cite{Nakanishi.Sneppen:1997}),
$D_a=d$ (assumed to hold on regular lattices by \cite{Ben-Hur.Biham:1996,
Chessa.etal:1999}, and in the present case confirmed for fractal lattices) and
$D_x(\tau_x - 1) = -\Sigma_x$
with
$\Sigma_a = \Sigma_s = \Sigma_t$ 
(narrow distribution assumption \cite{Jensen.etal:1989}). Using
$\tau = 2 - 2/D$,
$D_a = d$,
$\tau_a = 1+(D-2)/d$,
and
$\alpha = 1+(D-2)/z$, there are thus only
two functions to determine, which are best expressed in terms of $\epsilon=4-d$
since $d_c=4$ is the upper critical dimension \cite{Luebeck.Heger:2003a}, where
the exponents are known exactly. Writing $D=4-c_1^{(s)}\epsilon+c_2^{(s)}
\epsilon^2+\ldots$, at most two amplitudes $c_i^{(s)}$ can reasonably be
determined on the basis of the three data points available. A fit of $D$ with
only a linear term produces a very poor goodness-of-fit, which does not improve
satisfactorily by including a term quadratic in $\epsilon$. Omitting the
quadratic gives 
\begin{equation}\elabel{epsilon_expansion}
D=4-0.654(6)\epsilon+0.0079(10)\epsilon^3 
\end{equation}
with $q\approx0.095$
($c_1^{(s)}=0.60(4)$, $c_2^{(s)}=-0.05(3)$, $c_3^{(s)}=-0.019(7)$ with three
terms). Similarly, $z=2-0.239(4)\epsilon+0.0056(6)\epsilon^3$, however with
nearly vanishing goodness-of-fit.

\subsection{Fractal lattices}
Attempting to unify the above $\epsilon$-expansion obtained for regular lattices
with the results for fractals with Hausdorff dimension $d$ is bound to fail,
which is immediately clear when comparing the exponents found for the fractal
SITE lattice ($d=2$, \Tref{FRexponents}) with those for the regular
two-dimensional lattices, or the ARRO and the CRAB lattice, \Tref{FRexponents},
which have the same Hausdorff dimension. As is well understood, the basic scaling
relation \Eref{normal_diffusion} is valid only for regular lattices and has
to be generalised to
\begin{equation}
\elabel{anomalous_diffusion}
D (2 - \tau) = d_w
\end{equation}
with random walker dimension $d_w\ge2$ \cite{Avraham.Havlin:2000,
Huynh.etal:2010}.

It turns out, however, that $D$ is essentially a linear function of $d$
and $d_w$
\begin{equation}
\elabel{dwoverd_linear_Doverd}
D = a d + b d_w
\end{equation}
with \emph{the same coefficients $a$ and $b$ for both, regular and fractal
lattices}, which can be extracted from the $\epsilon$-expansion obtained
above with $d_c=4=2 d_w$, because $d_w=2$ on regular lattices, 
$a=c_1^{(s)}$ and $b=2(1-c_1^{(s)})$, so that on the basis of
\Eref{epsilon_expansion} $a=0.654(6)$ and $b=0.692(12)$ or $a=0.60(4)$ and
$b=0.80(8)$ depending
on $c_1^{(s)}$. Fitting the data in
\Tref{ADexponents} against \Eref{dwoverd_linear_Doverd} gives
$a=\qoflow{0.550(4)}$, $b=\qoflow{0.822(3)}$, and a fitting with the constraint
$b=2(1-a)$ from exactly known values of exponents at $d=d_c=4$ gives
$a=\qoflow{0.6061(5)}$,
$b=\qoflow{0.7878(10)}$.
\BFref{Doverd_dwoverd} shows $D/d$ as a function of $d_w/d$ for all
lattices listed in \Tref{ADexponents}. As expected from 
\Eref{dwoverd_linear_Doverd}, fractal and
regular lattices display essentially the same linear dependence 
$D/d=a+b d_w/d$. Above the upper critical
dimension $d_c=4$ regular lattices attain their mean-field values, 
$D=4$, $\tau=3/2$ \cite{Luebeck:2004} and $d_w=2$, therefore following
\Eref{dwoverd_linear_Doverd} with $a=0$ and $b=2$. One may wonder
whether fractal lattices with Hausdorff dimension greater than $d_c=4$
have correspondingly exponents $D=2 d_w$.

\begin{figure}
\includegraphics*[scale=0.35]{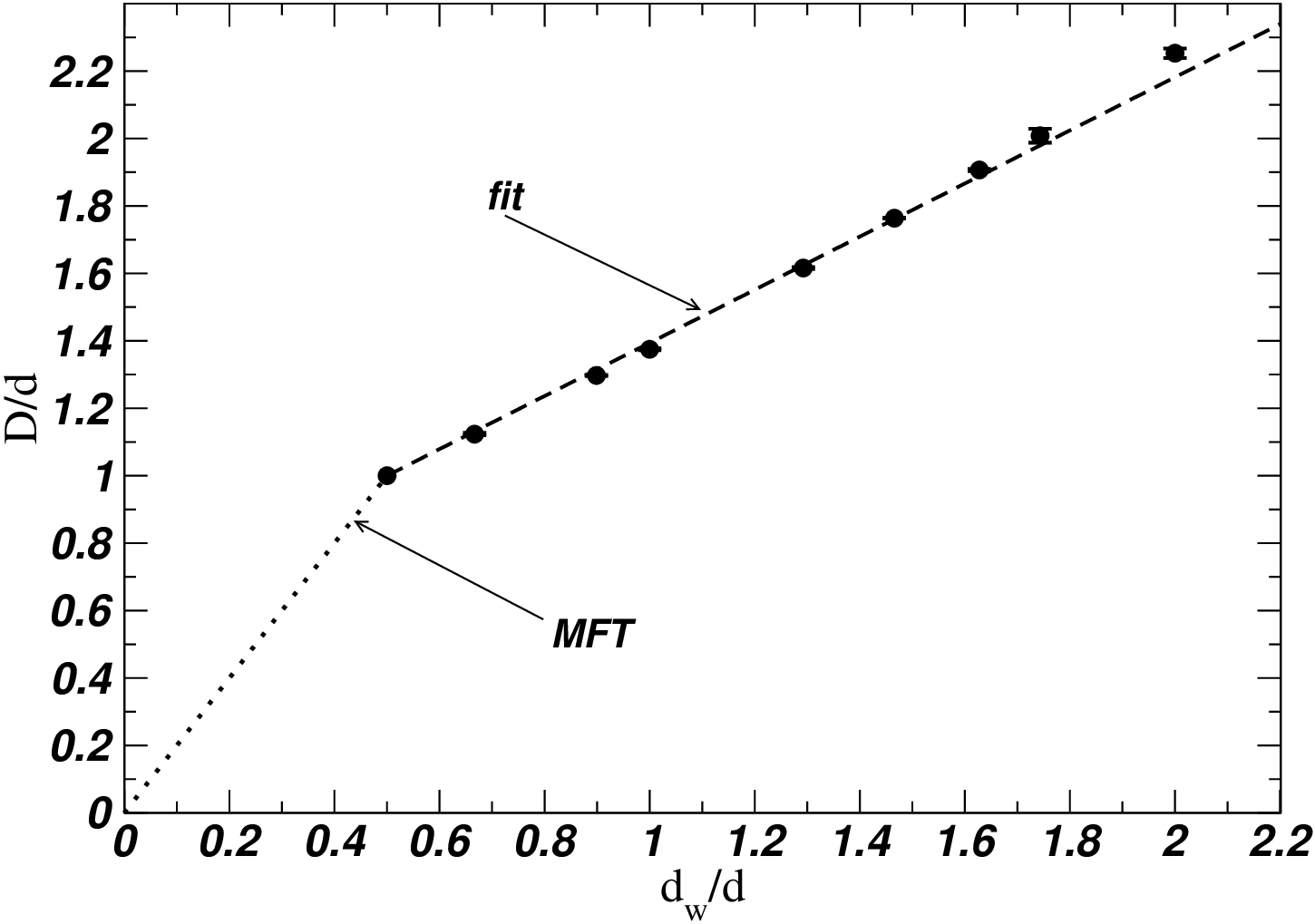}
\caption{\flabel{Doverd_dwoverd}The data of \Tref{ADexponents} plotted in the
form $D/d$ versus $d_w/d$ as suggested by \Eref{dwoverd_linear_Doverd}. The
dashed straight line is based on the estimates $a=\qoflow{0.6061(5)}$ and
$b=\qoflow{0.7878(10)}$, the dotted line is the mean field theory, $a=0$, $b=2$.
}
\end{figure}

The choice of rescaling exponent $D$ by dimension $d$ of the lattice is not 
random, but rather a natural choice, given that we perfomed all fitting
of $\mu_n^{(x)}$
against the number of sites $N$ rather than the lattices' linear length
$L$ and multiplied the results by the Hausdorff
dimension $d$. The gap exponents for the scaling in $N$ is $D/d$, which,
as it turns out, displays a very systematic dependence on $d_w/d$.

Further investigation shows that $D/d$ fits very well to
\begin{equation}
\elabel{Doverd_quadratic_tau}
\left(\frac{D}{d}\right)^2 (\tau - \tilde{a}) = \tilde{b}
\end{equation}
with $\tilde{a}=1.020(2)$ and $\tilde{b}=0.481(3)$ for \emph{all}
lattices which results in
\begin{equation}
D = 4 - 0.658(5) \epsilon + 0.00962(13) \epsilon^2 + 0.00161(3) \epsilon^3 +
\cdots
\end{equation}
using $D(2-\tau)=d_w=2$ for the regular ones.

The form of \Eref{Doverd_quadratic_tau} was
obtained by first fitting $\tau$ against $(D/d)^\kappa$, which gives
a $\kappa$ deviating from $-2$ by less than $2\%$.
The coefficient $\tilde{a}$ and $\tilde{b}$ are then fitted according
to \Eref{Doverd_quadratic_tau}.
\BFref{scaling_relation} compares that relation to results for lattices in
\emph{all} dimensions. In the same manner,  a similar relation can be obtained
for $z$ and $\alpha$,
\begin{equation}
\left(\frac{z}{d}\right)^\frac{3}{2} (\alpha-\tilde{a}) = \tilde{b}
\elabel{zoverd_twothirds_alpha}
\end{equation}
with $\qoflow{\tilde{a}=0.936(2)}$ and $\qoflow{\tilde{b}=0.3768(12)}$.

The above results suggest that the scaling in $N$ is more
suitable for fractals than the scaling in $L$. We suspect this is
related to $L$ 
not capturing the chemical distance, which is the
distance particles need to travel on the lattice, whereas $L$ is
measured as a Euclidean distance.
By using $d_w$, which is sensitive to the chemical distance, and
considering the scaling against $N$, 
which is a well-defined measure of the size for any lattice, we
are able to determine the relations above.

\begin{figure*}
\subfigure[Avalanche size exponents and fit, $D/d=(\tilde{b}/(\tau -
\tilde{a}))^{1/2}$]{
\includegraphics*[scale=0.3,angle=0]
{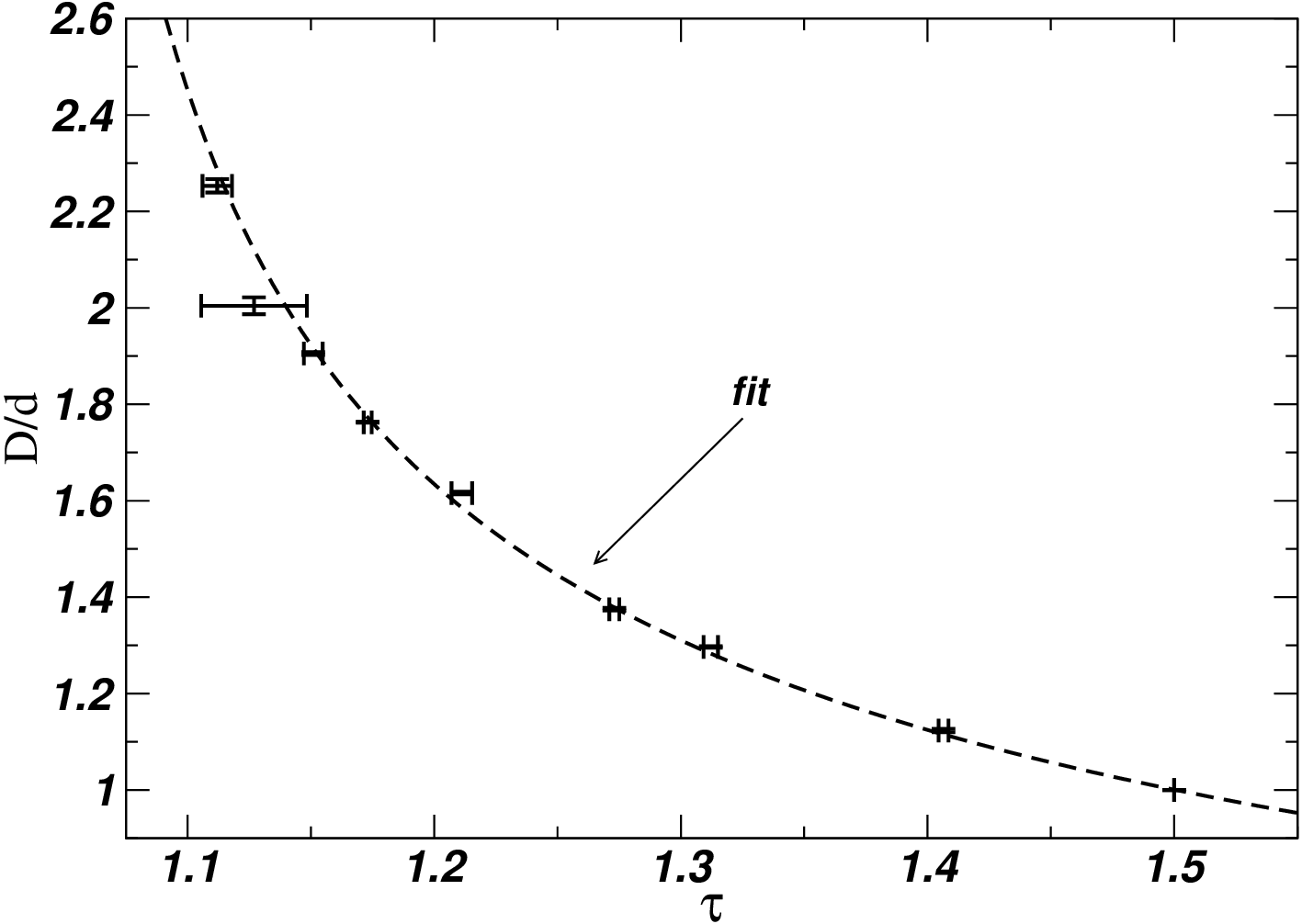}}
\subfigure[Avalanche duration exponents and fit, $z/d=(\tilde{b}/(\alpha -
\tilde{a}))^{2/3}$]
{\includegraphics*[scale=0.3,angle=0]{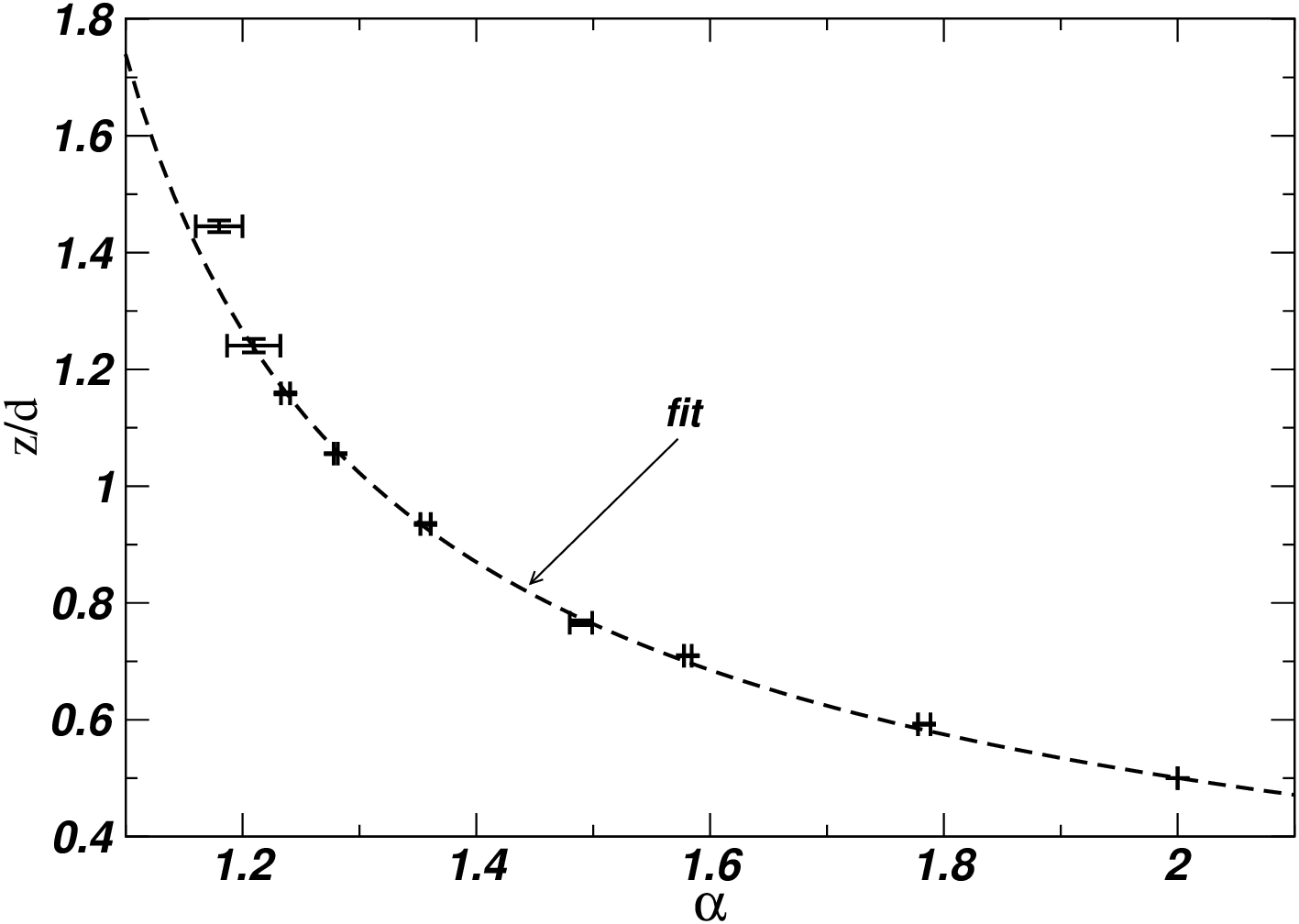}}
\caption{\flabel{scaling_relation}Fit of the exponents in all dimensions (on
regular and fractal lattices) against \Eref{Doverd_quadratic_tau}. The symbols
represent the data in \Tref{ADexponents}, the dashed lines are the fits as
described in the text, \Eref{Doverd_quadratic_tau} and
\Eref{zoverd_twothirds_alpha}, respectively.}
\end{figure*}

\section{Conclusion}
In conclusion, we studied Abelian Manna model on various three-dimensional and
fractal lattices with the aim to provide a complete picture about the model
below the upper critical dimension. The results confirm the consistent and
robust universal behaviour of the Manna model across different, regular
lattices, which allowed us to produce an $\epsilon$-expansion of avalanche
exponents below the upper critical dimension. 
A relation between critical exponents and lattice dimension is observed which
systematically reconciles integer dimensional with fractal lattices.

The authors are indebted to Andy Thomas, Dan Moore and Niall Adams for running
the SCAN computing facility at the Department of Mathematics of Imperial College
London. HNH thanks Lock Yue Chew for his support.

\bibliography{references}
\end{document}